\documentclass[journal]{IEEEtran}

\usepackage{booktabs}
\usepackage{xcolor}
\usepackage{colortbl}
\usepackage{amssymb}
\usepackage{algorithmic}
\usepackage{amsmath} 
\usepackage{amsfonts}
\newtheorem{theorem}{Theorem}

\newtheorem{definition}[theorem]{Definition}

\newtheorem{problem}[theorem]{Problem}

\ifCLASSINFOpdf
\usepackage[pdftex]{graphicx}
\else
\usepackage[dvips]{graphicx}
\fi
\usepackage{epstopdf}
\usepackage{array}
\ifCLASSOPTIONcompsoc
\usepackage[caption=false,font=normalsize,labelfont=sf,textfont=sf]{subfig}
\else
\usepackage[caption=false,font=footnotesize]{subfig}
\fi
\usepackage{float} 
\usepackage{fixltx2e}
\usepackage{color}
\usepackage{xcolor}
\usepackage{textcomp}
\usepackage[normalem]{ulem}
\usepackage{cite}
\usepackage{multibib} 
\usepackage{nomencl}
\makenomenclature
\usepackage{etoolbox}
\renewcommand\nomgroup[1]{%
	\item[\bfseries
	\ifstrequal{#1}{P}{Physical Variables}{%
		\ifstrequal{#1}{M}{Mathematical Symbols}{%
			\ifstrequal{#1}{S}{Subscripts and Superscripts}{}}} ]}

\usepackage{hyperref}
\hypersetup{
	colorlinks=true,
	linkcolor=blue,
	filecolor=magenta,      
	urlcolor=cyan,}
\usepackage{multirow}

\begin{document}
\title{Review on Set-Theoretic Methods for Safety Verification and Control of Power System}

\author{Yichen~Zhang,
	~Yan~Li,
	~Kevin~Tomsovic,
	~Seddik~M.~Djouadi,
    ~Meng~Yue
	\thanks{
		Y. Zhang is with Argonne National Laboratory, Lemont, IL 60439 USA (email: yichen.zhang@anl.gov).
		
		Y. Li is with Pennsylvania State University, State College, PA 16801, USA.
		
		K. Tomsovic and S. Djouadi are with The University of Tennessee, Knoxville, TN 37996, USA.
		
		M. Yue is with Brookhaven National Laboratory, Upton, NY 11973, USA.
		}}

\markboth{A\MakeLowercase{ccepted by} IET Energy System Integration \MakeLowercase{on} F\MakeLowercase{ebruary}, 2020 (DOI:)}%
{Shell \MakeLowercase{\textit{et al.}}: Bare Demo of IEEEtran.cls for IEEE Journals}
\maketitle

\begin{abstract}
	Increasing penetration of renewable energy introduces significant uncertainty into power systems. Traditional simulation-based verification methods may not be applicable due to the unknown-but-bounded feature of the uncertainty sets. Emerging set-theoretic methods have been intensively investigated to tackle this challenge. The paper comprehensively reviews these methods categorized by underlying mathematical principles, that is, set operation-based methods and passivity-based methods. Set operation-based methods are more computationally efficient, while passivity-based methods provide semi-analytical expression of reachable sets, which can be readily employed for control. Other features between different methods are also discussed and illustrated by numerical examples. A benchmark example is presented and solved by different methods to verify consistency.
\end{abstract}

\section{Introduction}
\begin{figure}[h]
	\centering
	\includegraphics[scale=0.17]{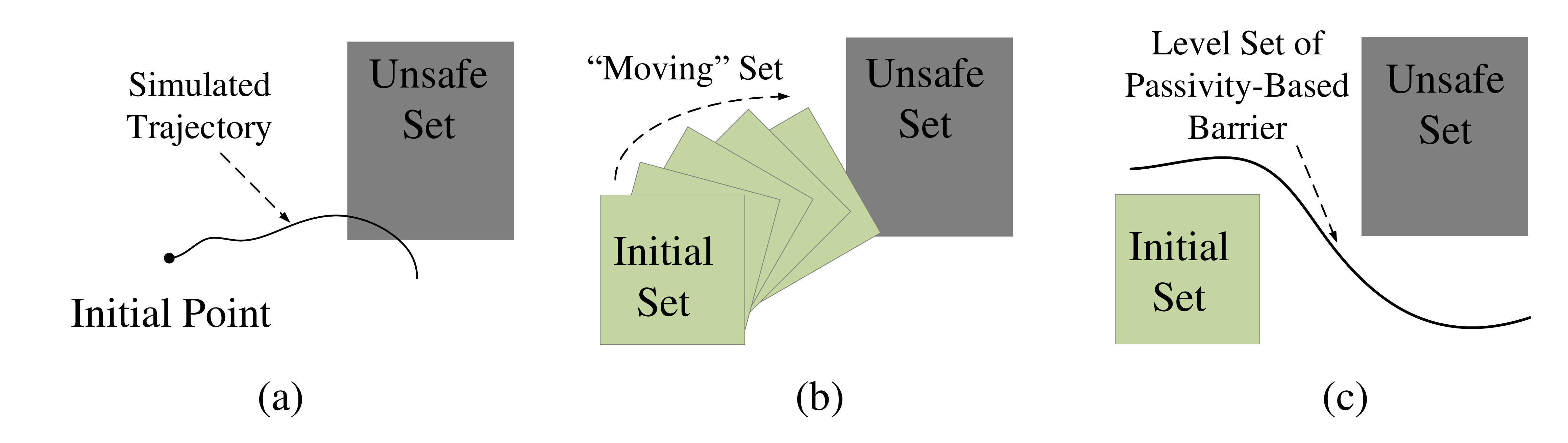}
	\caption{Safety verification based on (a) simulation, (b) set operating and (c) passivity.}
	\label{fig_Verifications}
\end{figure}

The importance of safety verification increases tremendously for modern engineering systems whose functions are safety-critical such as the transportation systems and power systems. Safety verification is to secure the evolution of dynamic system states, or more specifically to prove that there exists no trajectory entering a set of forbidden, normally denoted as unsafe states \cite{althoff2010reachability}. Most safety verification approaches can be categorized into three main groups: simulation, set operation and passivity-based methods, which are illustrated in Fig. \ref{fig_Verifications}. The traditional and most widely-used method is simulation. When the system is subjected to input and parameter uncertainties, sampling over the sets is a premise of the simulation task, which requires the statistical information. The objective is to generate a finite set of trajectories that will exhibit all the behaviors of the system \cite{le2009reachability}, or provide a sufficient confidence level \cite{clarke2011statistical}. Rapidly-exploring random trees \cite{Bhatia2004}, robust test case generation\cite{julius2007robust}, and Monte Carlo simulation \cite{hegazy2003adequacy} are the major techniques to achieve this goal. In power industry, this procedure is a routine known as the dynamic security assessment (DSA) \cite{Ni2002} and extremely important to guarantee a reliable electric energy transmission. 

Although the simulation method is efficient, it cannot handle the uncertainties with only unknown-but-bounded assumption. More importantly, simulation is often terminated inconclusively if no counter-example is produced, since there exist infinitely many possible trajectories \cite{althoff2010reachability}. 
Set-theoretic methods can be employed to tackle these issues. Set-theoretic methods can be loosely defined as any method which exploits the properties of the properly chosen sets or constructed sets in the state space \cite{blanchini2008set,villanueva2015set}. The set operation-based methods aim to evaluate the bounds of all possible trajectories at each time step in an over-approximated fashion. The bounds can be obtained by solving nonlinear optimization \cite{choi2016propagating}, interval mathematics \cite{althoff2007reachability, Althoff2014}, or the Hamilton-Jacobi partial differential equations \cite{Tomlin2003,Jin2010}. Similar to the simulation, these methods also rely on numerical discretization of the continuous systems as well as the explicit representation of the system solutions. Therefore, although provable bounds can be obtained, the computation is intensive and the results may be conservative to a certain level. 

On the other hand, the passivity-based methods search for certificates that prove the safety of the system. A common technique is to compute a function in terms of the system states, the zero level set of which provides a "barrier" between the possible system trajectories and the given unsafe region, in the sense that no trajectory of the system starting from the initial set can cross this level set to reach the unsafe region \cite{Prajna2007a}. It is, in spirit, closely linked to the Lyapunov theory. The Lie deviation is used to represent the underlying vector field, and thus no explicit solution needs to be computed. The invariance principle can guarantee the safety of a system over an infinite time horizon. Since the function is in terms of the system states, it can naturally provide a supervisory function if the state estimation is available \cite{zhang2018set}, admitting an extension to hybrid systems \cite{zyc_hybrid_JCS_2017}. Nevertheless, the condition is only sufficient. The certificate searching algorithms can terminate inconclusively.

In power networks, with deep penetration of converter-interfaced devices, such as different types of renewable energy, electric vehicles, flexible alternating current transmission systems (FACTS) and high-voltage direct current (HVDC) electric power transmission systems, uncertainty sources continue increasing. The traditional simulation and DSA suffer from a combinatorial explosion and lack of statistical information. The set-theoretic methods are appealing as alternative solutions. In this paper, we will review different both the set-theoretic and passivity-based methods in the categorized manner as well as their applications in power systems. All reviewed techniques with their application in power systems are concluded in Table \ref{tab_conclusion}.

The outline of the paper is as follows. In Section \ref{sec_reach_set}, the set operation-based methods, including Lagrangian and Eulerian methods, are reviewed. In Section \ref{sec_reach_BC}, the passivity-based methods are presented, where different algorithmic solutions are discussed with an illustration of several examples, followed by the conclusions in \ref{sec_con}.

\subsubsection{Preliminaries and Notations}
Safety denotes the property that all system trajectories stay within the given bounded regions, thus, the equipment damage or relay triggering can be avoided. Note this is similar, but not identical, to what is called the security in power industry but for the purposes of this paper we will assume satisfying safety conditions ensures a secure operation. Consider the dynamics of a power system governed by a set of ordinary differential equations (ODEs) as
\begin{equation}
\label{eq_ode}
\dot{x}(t)=f(x(t),d(t)),\quad t\in[0,T]
\end{equation}
where $T>0$ is a terminal time, $x(\cdot): [0,T]\rightarrow\mathbb{R}^{n}$ denotes the vector of state variables and $d(\cdot): [0,T]\rightarrow\mathbb{R}^{m}$ denotes the vector of certain disturbances, such as, generation losses or abrupt load changes. The vector fields $f: \mathbb{R}^{n}\times\mathbb{R}^{m}\rightarrow\mathbb{R}^{n}$ is such that for any $d$ and initial condition $x_0$, the state equation (\ref{eq_ode}) has a unique solution defined for all $t\in[0,T]$, denoted by $\phi(t;d(t),x_0): [0,T]\rightarrow\mathbb{R}^{n}$. Note that we employ a semicolon to distinguish the arguments and the trajectory parameters.

For the verification tasks in power systems, the disturbances may be assumed bounded in the set $D\subseteq\mathbb{R}^{m}$, that is, $d(\cdot): [0,T]\rightarrow D$. Let $X\subseteq\mathbb{R}^{n}$ be the computational domain of interests, $X_{I}\subseteq X$ be the initial set and $X_{U}\subseteq X$ be the unsafe set, then the formal definition of the safety property is given as follows.
\begin{definition}[Safety]
	\label{thm_safety_def}
	Given (\ref{eq_ode}), $X$, $X_{I}$, $X_{U}$ and $D$, the \emph{safety} property holds if there exists no time instant $T\geq 0$ and no piece-wise continuous and
	bounded disturbance $d: [0,T]\rightarrow D$ such that $\phi(t;d(t),x_{0})\cap X_{U}\neq\varnothing$ for all $t\in[0,T]$ and $x_0\in X_{I}$.
\end{definition}

\section{Set Operation-Based Methods}\label{sec_reach_set}
\begin{figure}[h]
	\centering
	\includegraphics[scale=0.17]{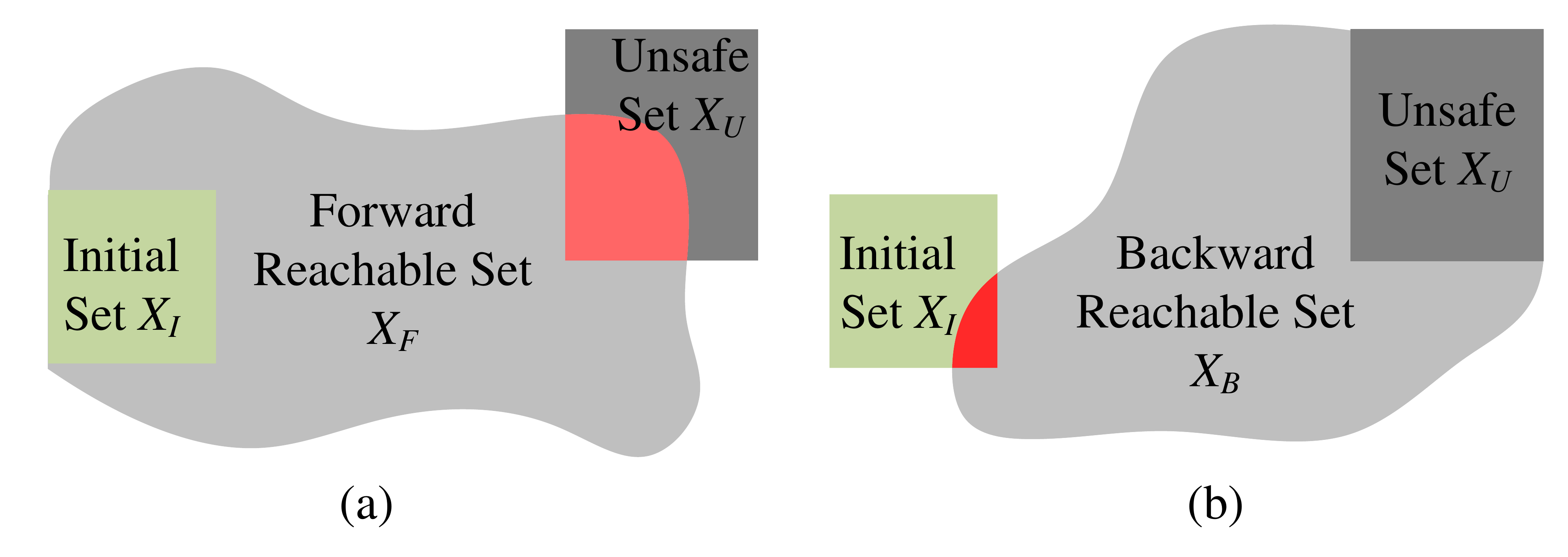}
	\caption{Safety verification based on (a) forward reachable set, (b) backward reachable set.}
	\label{fig_ReachVerifications}
\end{figure}

The set operation-based verification can be categorized in different ways. From the \emph{execution} point of view, the set operation-based verification can be conducted using either the forward reachable sets or backward reachable sets as illustrated in Fig. \ref{fig_ReachVerifications} \cite{Maidens2013}. In the forward verification, the reachable set for the given initial set denoted by $X_{F}$ is computed under the system vector fields to examine whether $X_{F}$ intersects with $X_{U}$. While, in the backward verification, the reachable set denoted by $X_{B}$ is computed in the reverse time and the intersecting condition between $X_{I}$ and $X_B$ is examined.

From the \emph{computation} point of view, there are Lagrangian and Eulerian methods \cite{Maidens2013}. Both types of methods can be executed in either the forward or the backward setting. Lagrangian methods work with linear systems and seek efficient over-approximation of the reachable sets. Eulerian method (also known as the level set method), which can deal with the general dynamic systems, is to calculate as closely as possible the true reachable set by computing a numerical solution to the Hamilton-Jacobi partial differential equation (HJ PDE). Both methods are briefly introduced in this subsection.

\subsection{Lagrangian Methods}\label{sec_sub_reach_set_L}
Lagrangian methods compute over-approximation of the reachable sets by propagating the sets under the vector fields of linear systems efficiently. The efficiency relies on the special representations of sets as boxes, ellipsoids, polytopes, support functions and so on. Among all representations, the ellipsoids \cite{Kurzhanskiy2007} and zonotopes \cite{Girard2005}, a sub-class of polytopes, are widely-used. It is worth mentioning that nonlinear differential-algebraic systems have been addressed in \cite{Althoff2014} by using the conservative linearization.

\begin{table*}[htbp!]
	\caption{Application of Set Theoretic Methods in Power and Energy Systems}
	\label{tab_conclusion}
	\begin{center}
		\begin{tabular}{ | m{1cm} | m{1.5cm}| m{3.6cm} | m{3.5cm} | m{6cm} | }
			\hline
			Category & Technique & Advantage & Disadvantage & Topics and References\\
			\hline
			\multirow{7}{4em}{Set operation} & \multirow{7}{4em}{Lagrangian-Ellipsoid} &  & Only applicable to linear systems (nonlinear systems need linearization) & \cite{Jiang2013}\cite{YuChristineChen2011} Uncertainty impact on power flow \\
			& &Leading to a convex optimization & & \cite{Chen2012} Uncertainty impact on dynamic performance \\
			& & & No closed-form description & \cite{Hope2011} Large-signal behavior of DC-DC converters \\
			& & & & \cite{Xu2016} Locational impacts of virtual inertia on the frequency responses\\
			& & & Shape limitation induced conservatism & \\
			\hline
			\multirow{7}{4em}{Set operation} & \multirow{7}{4em}{Lagrangian-Zonotope} &  &  & \cite{Pico2013} Frequency dynamics with HVAC and HVDC transmission lines \\
			& & Flexible computation complexity based on preference setting & Only applicable to linear systems (nonlinear systems need linearization) & \cite{Pico2014b}\cite{Pico2014} Voltage ride-through capability of wind turbine generators \\
			& & & & \cite{Jiang2014} Uncertainty impact on power flow \\
			& & Adjustable shape and representation power & No closed-form description & \cite{Althoff2014}\cite{Althoff2014b}\cite{El-Guindy2017} Transient stability  \\
			& & & & \cite{El-Guindy2016} Load-following capabilities maximization \\
			& & & Underlying interval analysis is conservative after many steps & \cite{Al-Digs2016} Feasible nodal power injections estimation \\
			& & & & \cite{li2017formal,li2018networked,li2018distributed} Microgrid stability \\
			\hline
			\multirow{5}{4em}{Set operation} & \multirow{5}{4em}{Lagrangian-Supporting function} &  &  & \\
			& & Sometimes rendering to efficient optimization problems & Only applicable to linear systems (nonlinear systems need linearization) & \cite{wang2019reachability} Power electronic system \\
			& & &  &  \\
			& & & No closed-form description &  \\
			& & &  &  \\
			\hline
			\multirow{4}{4em}{Set operation} & \multirow{4}{4em}{Eulerian} & High accuracy & \multirow{4}{12em}{Extremely high computation complexity (computational feasible up to 4th-order systems)} & \\
			& & & & \cite{Jin2010}\cite{Susuki2012} Transient stability \\
			& & Applicable to nonlinear systems & & \cite{Susuki2007} Voltage stability \\
			& & & &  \\
			\hline
			\multirow{6}{4em}{Passivity} & \multirow{6}{4em}{Sums of square representation} &  &  & \cite{Wisniewski2013}\cite{Laurijsse2014a} Supervisory control for emergency wind turbines shutdown \\
			& & Flexible computation complexity
			based on polynomial order choice & Trade-off between representation accuracy and computation complexity & \cite{Pedersen2016} Voltage constraint satisfaction \\
			& & & & \cite{zhang2018set}\cite{zyc_hybrid_JCS_2017} Supervisory control for grid supportive functions\\
			& & Close-form solution and readily applicable for control & Leading to large-scale SDP & \cite{kundu2019distributed} Voltage compensation control in inverter-based microgrids \\
			& & & Only admitting sufficient condition  & \cite{anghel2013algorithmic, kundu2015stability, mishra2017stability, mishra2019transient, josz2019transient} Lyapunov function for transient stability analysis \\
			& & & &  \\
			\hline
			\multirow{2}{4em}{Passivity} & \multirow{2}{4em}{Linear representation} & Leading to LP & Shape limitation induced conservatism & \multirow{2}{12em}{\emph{currently no application in power systems}}\\
			& & Close-form solution and readily applicable for control & Only admitting sufficient condition & \\
			\hline
		\end{tabular}
	\end{center}
\end{table*}\par


The essence of the Lagrangian methods is to find the boundary of all possible trajectories of a nonlinear differential-algebraic system under various input and parameter uncertainties \cite{althoff2019reachability}. Specifically, through the Lagrangian methods, one can compute the reachable sets for each short time interval $\eta_j=[t_j, t_{j+1}]$, where $t_j$ and $t_{j+1}$ are time steps. 

For instance, when the system is modeled by using a set of differential-algebraic equations as shown in (\ref{Eq_DAE}), the state matrix $\boldsymbol{A}$ can be obtained through $\boldsymbol{f}_{\boldsymbol{x}} -\boldsymbol{f}_{\boldsymbol{y}} \boldsymbol{g}_{\boldsymbol{y}}^{-1} \boldsymbol{g}_{\boldsymbol{x}}$, where  $\boldsymbol{f}_{\boldsymbol{x}}={\partial \boldsymbol{f}}/{\partial \boldsymbol{x}}$ is the partial derivative matrix of differential equations with respect to state variables, $\boldsymbol{f}_{\boldsymbol{y}}={\partial \boldsymbol{f}}/{\partial \boldsymbol{y}}$ is the partial derivative matrix of differential equations with respect to the algebraic variables, $\boldsymbol{g}_{\boldsymbol{x}}={\partial \boldsymbol{g}}/{\partial \boldsymbol{x}}$ is the partial derivative matrix of algebraic equations with respect to the state variables, and $\boldsymbol{g}_{\boldsymbol{y}}={\partial \boldsymbol{g}}/{\partial \boldsymbol{y}}$ is the partial derivative matrix of algebraic equations with respect to the algebraic variables.

\begin{equation}  \label{Eq_DAE}
\left \{	\begin{aligned}
&\dot{\boldsymbol{x}}(t)=\boldsymbol{f}\big(\boldsymbol{x}(t),\boldsymbol{y}(t),\boldsymbol{d}(t)\big)\\
&\boldsymbol{0}=\boldsymbol{g}\big(\boldsymbol{x}(t),\boldsymbol{y}(t),\boldsymbol{d}(t)\big), \quad t\in[0,T]
\end{aligned} \right.
\end{equation}

\begin{figure}[h]
	\centering
	\includegraphics[scale=0.7]{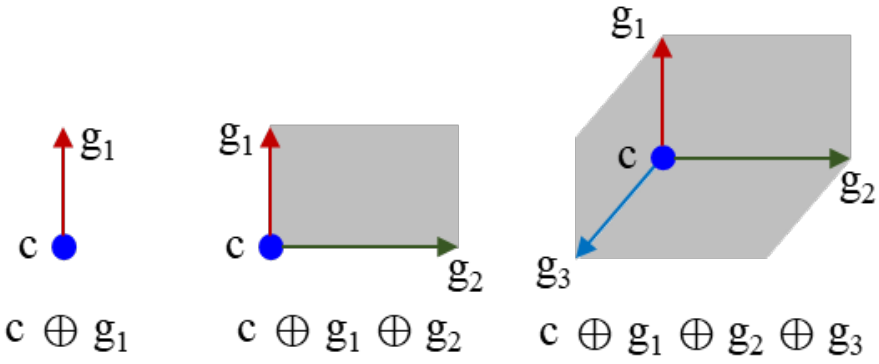}
	\caption{Illustration of one-, two- and three-dimension zonotope.}
	\label{fig_Zonotope}
\end{figure}

One important step for reachable set calculation is to properly model the uncertainties $\boldsymbol{d}(t)$. Although the uncertainties in the power grid are time-varying, the most frequent uncertainties and their ranges can be obtained through the measurements. Taking into account the dependence between uncertainties, instead of modeling those uncertainties one by one, which is inefficient, a sub-class of polytopes are widely-used. Taking zonotope as an example, Fig. \ref{fig_Zonotope} illustrates the system uncertainties by using one-, two- and three-dimensional zonotopes. Mathematically, a zonotope $\boldsymbol{d}(t)$ can be modeled by a center and multiple generators as follows~\cite{althoff2014formal,althoff2011zonotope}:
\begin{equation} \label{Eq21}
\boldsymbol{d}(t)=\{\boldsymbol{c}+\sum_{i=1}^m\alpha_i\boldsymbol{g}_i\mid\alpha_i\in [-1,1]\},
\end{equation}
where $\boldsymbol{c} \in \mathbb{R}^n$ is the center and $\boldsymbol{g}_i \in \mathbb{R}^n$ are generators. 

Besides the regular zonotope, several other polytopes can be adopted according to the features of uncertainties, e.g., using a sparse polynomial zonotope method \cite{kochdumper2019sparse} to model the interdependence among uncertainties.

After obtaining the system state matrix $\boldsymbol{A}$ and properly modeling the uncertainties $\boldsymbol{d}(t)$, the reachable sets at each time step and during time steps can be over-approximated via the following closed-form solutions: 
\begin{equation}\label{Eq_reachable_set}
\mathcal{S}(t_{j+1})=\mathrm{e}^{\boldsymbol{A}\eta_j}\mathcal{S}(t_j)\oplus\phi_0(\boldsymbol{A}, \eta_j, \boldsymbol{\mathcal{Z}_0})\oplus \varphi_\Delta(\boldsymbol{\mathcal{Z}_{\Delta}},\eta_j),  
\end{equation}
\begin{flalign}\label{Eq_reachable_set_2}	
\mathcal{S}(\eta_j) &=C(\mathcal{S}(t_j), \mathrm{e}^{\boldsymbol{A}\eta_j} \mathcal{S}(t_j)\oplus\phi_0(\boldsymbol{A}, \eta_j, \boldsymbol{\mathcal{Z}_0})) 	 \notag\\
& \quad \oplus \varphi_\Delta(\boldsymbol{\mathcal{Z}_{\Delta}},\eta_j)\oplus \psi, 
\end{flalign}

\noindent where $\mathcal{S}(t_{j+1})$ is the reachable set at the time step $t_{j+1}$; $\mathcal{S}(\eta_j)$ is the reachable set during time step $t_j$ and $t_{j+1}$; $\mathrm{e}^{\boldsymbol{A}\eta_j}\mathcal{S}(t_j)$ is the impact of the history reachable set on the current one; $\phi_0(\boldsymbol{A}, \eta_j, \boldsymbol{\mathcal{Z}_0})$ represents the increment of reachable set caused by the deterministic uncertainty $\boldsymbol{\mathcal{Z}_0}$ (the center of the zonotope); $\varphi_\Delta(\boldsymbol{\mathcal{Z}_{\Delta}},\eta_j)$ represents the increment of reachable set caused by the uncertainty $\boldsymbol{\mathcal{Z}_{\Delta}}$; $\psi$ represents the increment of the reachable set caused by the curvature of trajectories from $t_j$ to $t_{j+1}$; $C(\cdot)$ means the convex hull calculation; and $\oplus$ means Minkowski addition. The items involved in (\ref{Eq_reachable_set}) and (\ref{Eq_reachable_set_2}) can be further expressed as follows~\cite{althoff2008reachability,althoff2014formal, schurmann2018reachset}:

\begin{equation}\label{Eq_phi}
\phi_0(\boldsymbol{A}, \eta_j, \boldsymbol{\mathcal{Z}_0}) =\bigg\{\sum\limits_{i=0}^{\beta} \frac{\boldsymbol{A}^{i}\eta_j^{i+1}}{(i+1)!}\oplus \mathcal{F} \bigg\}\boldsymbol{\mathcal{Z}_0},
\end{equation}
\begin{equation}\label{Eq_f}
\mathcal{F}= \big[-\Upsilon(\boldsymbol{A},\eta_j)\eta_j, \Upsilon(\boldsymbol{A},\eta_j)\eta_j\big],
\end{equation}
\begin{flalign}
\varphi_\Delta(\boldsymbol{\mathcal{Z}_{\Delta}},\eta_j) &=\sum\limits_{i=0}^{\beta} \bigg(\frac{\boldsymbol{A}^{i}\eta_j^{i+1}}{(i+1)!} \boldsymbol{\mathcal{Z}_{\Delta}}\bigg) \oplus\big\{\mathcal{F} \cdot\boldsymbol{\mathcal{Z}_{\Delta}}  \big\},   
\end{flalign}
\begin{equation}
\psi= \big\{\big(\mathcal{I}\oplus \mathcal{G}\big)\cdot\mathcal{S}(t_{j})\big\}
\oplus \big\{\big(\tilde{\mathcal{I}}\oplus \mathcal{F} \big)\cdot\boldsymbol{\mathcal{Z}_0} \big\},
\end{equation}
\begin{flalign}\label{Eq_g}
\mathcal{G}=  [-\Upsilon(\boldsymbol{A},\eta_j), \Upsilon(\boldsymbol{A},\eta_j)].
\end{flalign}

And $\Upsilon(\boldsymbol{A},\eta_j)$, $\mathcal{I}$, $\tilde{\mathcal{I}}$ involved in (\ref{Eq_f})-(\ref{Eq_g}) are given as follows:
\begin{equation}
\Upsilon(\boldsymbol{A},\eta_j)=e^{|\boldsymbol{A}|\eta_j}-\sum\limits_{i=0}^{\beta} \frac{(|\boldsymbol{A}|\eta_j)^{i}}{i!},
\end{equation}
\begin{equation}
\mathcal{I}=\sum\limits_{i=2}^{\beta} \big[(i^{\frac{-i}{i-1}}-i^{\frac{-1}{i-1}})\eta_j^i,0\big] \frac{\boldsymbol{A}^{i}}{i!},
\end{equation}
\begin{equation}
\tilde{\mathcal{I}}=\sum\limits_{i=2}^{\beta+1} \big[(i^{\frac{-i}{i-1}}-i^{\frac{-1}{i-1}})\eta_j^i,0\big] \frac{\boldsymbol{A}^{i-1}}{i!}.
\end{equation}

\begin{figure}[h]
	\centering
	\includegraphics[scale=0.8]{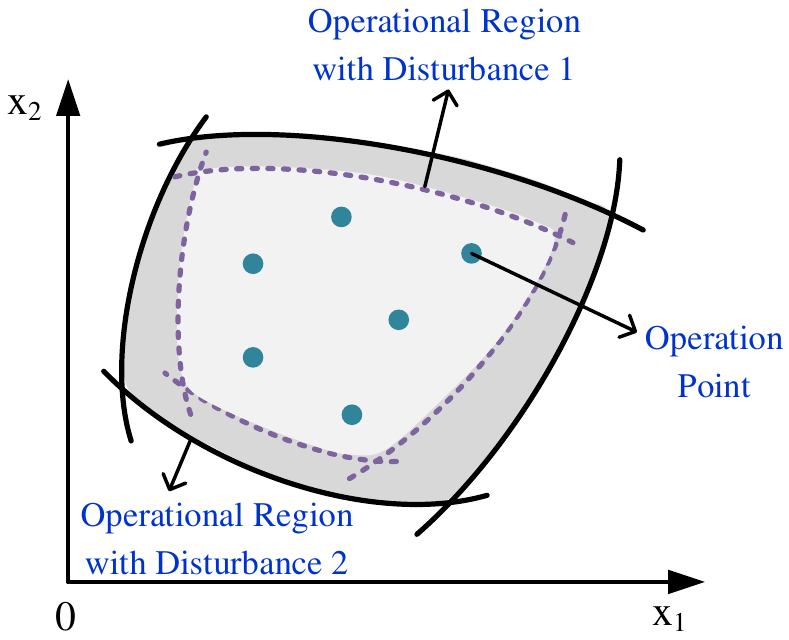}
	\caption{Illustration of system's operational region under disturbances.}
	\label{fig_margin}
\end{figure}

Overall, (\ref{Eq_reachable_set}) and (\ref{Eq_reachable_set_2}) show the reachable sets calculation over time through the centralized Lagrangian methods with the computational complexity of  $O(n^5)$. This method can be used for control verification~\cite{el2017formal,susuki2008verifying, althoff2019effortless, koschi2019computationally}, identifications of stability regions~\cite{li2017formal}, transient stability analysis~\cite{el2017compositional}, model conformance~\cite{kochdumper2020establishing, roehm2019model}, risk evaluation~\cite{jin2019risk}, etc. For instance, \cite{althoff2014reachability} computes reachable sets of nonlinear differential-algebraic systems under uncertain initial states and inputs. It can be further developed and used for control verification of power system properties. A quasi-diagonalized Geršgorin theory was established in \cite{li2017formal} and then combined with the centralized Lagrangian method to efficiently identify microgrids' stability region under disturbances as illustrated in Fig. \ref{fig_margin}. It shows the impact of disturbances on a networked microgrid system's stability margin.

The centralized Lagrangian methods have many applications in power system. For instance, they can be used for power system forecast and monitoring, verification of new control or dispatch strategies, identification of critical disturbances or scenarios, etc.

Although the centralized Lagrangian methods are powerful in evaluating system dynamics subject to disturbances, it is computationally impractical to apply these methods to a large-scale nonlinear dynamic system due to the high dimensionality and operational flexibility \cite{li2018networked}. A distributed formal analysis~\cite{althoff2014formal,el2017compositional} (or compositional formal analysis) is studied for efficient calculation and verification. ~\cite{althoff2014formal} abstracts the dynamics of a large-scale system to linear differential inclusions by using the full model and then compositionally computes the set of linearization errors. \cite{el2017compositional} splits a large-scale interconnected grid into subsystems for which the reachable sets are computed separately.

\subsection{Eulerian Methods}\label{sec_sub_reach_set_E}
\emph{Strictly speaking}, the Eulerian method is known as the level set method. In this method, the initial set at time $t$ is implicitly represented by the zero sublevel sets of an appropriate function denoted by $\phi(x,t): \mathbb{R}^{n}\times\mathbb{R}\rightarrow\mathbb{R}$, where the surface of the initial set at time $t$ is expressed as  $\phi(x,t)=0$. Consider a small variation along this surface, i.e., moving $(x,t)$ to a neighboring point $(x+dx,t+dt)$ on the surface, the variation in $\phi$ will be zero
\begin{align}\label{eq_surface}
d\phi=\phi(x+dx,t+dt)-\phi(x,t)=0
\end{align}
which finally leads to the HJ PDE
\begin{align}
\label{eq_HJ_PDE}
\sum\limits_{i}\frac{\partial\phi}{\partial x_{i}}\frac{dx}{dt}+\frac{\partial\phi}{\partial t}=0
\end{align}
The state evolution is governed by the ODE in (\ref{eq_ode}). Thus, Eq. (\ref{eq_HJ_PDE}) is cast as follows
\begin{align}
\label{eq_HJ_PDE_f}
\sum\limits_{i}\frac{\partial\phi}{\partial x_{i}}f(x,d)+\frac{\partial\phi}{\partial t}=0
\end{align}
This PDE describes the propagation of the reachable set boundary as a function of time under the system vector field. By solving the PDE, the precise reachable sets can be obtained, and therefore this method is known as the convergent approximation \cite{Tomlin2003}. Transient stability \cite{Jin2010}\cite{Susuki2012} and voltage stability \cite{Susuki2007} are analyzed using this approach. However, to obtain numerical solutions, one needs to discretize the state space, which leads to an exponentially increasing computational complexity and limits its applications to systems with no more than four continuous states \cite{Althoff2014}.\par

\emph{Broadly speaking}, the initial set at time $t$ can be expressed alternatively like the occupation measure in \cite{Henrion2014}. Propagating such a measure (set-valued function) will lead to the Liouville's PDE. In spirit, the type of methods is closer to the level set method, although may be in a different category from the computation perspective.

\section{Passivity-Based Methods}\label{sec_reach_BC}
Different from the set operation-based approaches, which can be regarded essentially as the set-valued simulation, the passivity-based methods exploit and extract invariance features from the vector field of (\ref{eq_ode}), and provide certificates (as a function of system states and thus in state space) proving \emph{unreachability} to unsafe sets. Such a certificate is denoted as a barrier certificate \cite{Prajna2007a}. If these unsafe sets are infinitely far from the system's equilibrium point(s), then the certificate provides a stability proof, and are therefore a Lyapunov function. Essentially the barrier certificates and the Lyapunov functions are the same. The key to computing a barrier certificate is to search the functions that are point-wise positive over a set. In this section, the barrier certificate and its extension will be discussed first. Then, theorems and algorithms that admit the positivity condition are introduced, followed by a review of the barrier certificate applications in power systems.

\subsection{Barrier Certificate and Region of Safety}
The concept of the barrier certificate for safety verification is firstly proposed in \cite{Prajna2007a} and formally stated in the following theorem.
\begin{theorem}
	\label{thm_barrier_a}
	Let the system $\dot{x}=f(x,d)$, and the sets $X\subseteq\mathbb{R}^{n}$, $X_{I}\subseteq X$, $X_{U}\subseteq X$ and $D\in\mathbb{R}^{m}$ be given, with $f\in C(\mathbb{R}^{n+m},\mathbb{R}^{n})$. If there exists a differentiable function $B:\mathbb{R}^{n}\rightarrow \mathbb{R}$ such that
	\begin{subequations}
		\begin{align}
		B(x)\leq 0& \qquad \forall x \in X_{I}\label{eq_barrier_a1}\\
		B(x)> 0& \qquad \forall x \in X_{U}\label{eq_barrier_a2}\\
		\dfrac{\partial B(x)}{\partial x}f(x,d)<0& \qquad \forall (x,d) \in X\times D  \label{eq_barrier_a3}
		\end{align}
	\end{subequations}
	then the safety of the system in the sense of Definition \ref{thm_safety_def} is guaranteed.
\end{theorem}

The function $B(x)$ satisfied the above theorem is called a barrier certificate. The zero level set of $B(x)$ defines an invariant set containing $X_{I}$, that is, no trajectory starting in $X_{I}$ can cross the boundary to reach the unsafe set. It is guaranteed by the negativity of $B(x)$ over $X_{I}$ and the decrease of $B(x)$ along the system vector fields. Although conditions in Theorem \ref{thm_barrier_a} is convex, it is rather conservative due to the satisfaction of (\ref{eq_barrier_a3}) over the entire state space. A non-convex but less conservative condition is also proposed in \cite{Prajna2007a} as follows.
\begin{theorem}
	\label{thm_barrier_b}
	Let the system $\dot{x}=f(x,d)$, and the sets $X\subseteq\mathbb{R}^{n}$, $X_{I}\subseteq X$, $X_{U}\subseteq X$ and $D\in\mathbb{R}^{m}$ be given, with $f\in C(\mathbb{R}^{n+m},\mathbb{R}^{m})$. If there exists a differentiable function $B:\mathbb{R}^{n}\rightarrow \mathbb{R}$ such that
	\begin{subequations}
		\begin{align}
		B(x)\leq 0& \qquad \forall x \in X_{I}\label{eq_barrier_1}\\
		B(x)> 0& \qquad \forall x \in X_{U}\label{eq_barrier_2}\\
		\dfrac{\partial B}{\partial x}f(x,d)<0& \quad \forall (x,d) \in X\times D \quad \mathrm{s.t.} \quad B(x)=0 \label{eq_barrier_3}
		\end{align}
	\end{subequations}
	then the safety of the system in the sense of Definition \ref{thm_safety_def} is guaranteed.
\end{theorem}

Eq. (\ref{eq_barrier_3}) reduces conservatism in the sense that the passivity condition only needs to hold on the zero level set of $B(x)$ instead of the whole state space. Compositional barrier certificates are discussed in \cite{Sloth2012} and \cite{Sloth2012a} for verification of the interconnected systems. By using the barrier certificate, safety can be verified without explicitly computing trajectories nor reachable sets.

In the above methods, the initial condition $X_{I}$ has to be known. In many problems, however, we would like to know the set of initial condition that only admits safe trajectories. Analogous to the region of attraction in describing stability features, the concept \emph{region of safety} is proposed in \cite{zyc_hybrid_JCS_2017}. In addition, estimation of the largest region of safety (ROS) will be important to controller synthesis. The corresponding conceptual problem is proposed in \cite{zyc_hybrid_JCS_2017}, and formally formulated as below.
\begin{problem}
	\label{thm_max_volume}
	Let $\dot{x}=f(x,d)$, $X$, $X_{U}$ and $D$ be given. The region of safety $X_{I}$ is obtained by solving:
	\begin{align*}
	&\max_{X_{I},B(x)}& &\quad \text{Volume}(X_{I}) \\
	& \text{s. t. }& &B(x)\leq 0  \quad \forall x \in X_{I} \\
	& & &B(x)> 0  \quad \forall x \in X_{U} \\
	& & &\dfrac{\partial B}{\partial x}f(x,d)<0\quad\forall (x,d) \in X\times D \text{ s.t. } B(x)=0
	\end{align*}
\end{problem}\par

\begin{figure}[h]
	\centering
	\includegraphics[scale=0.25]{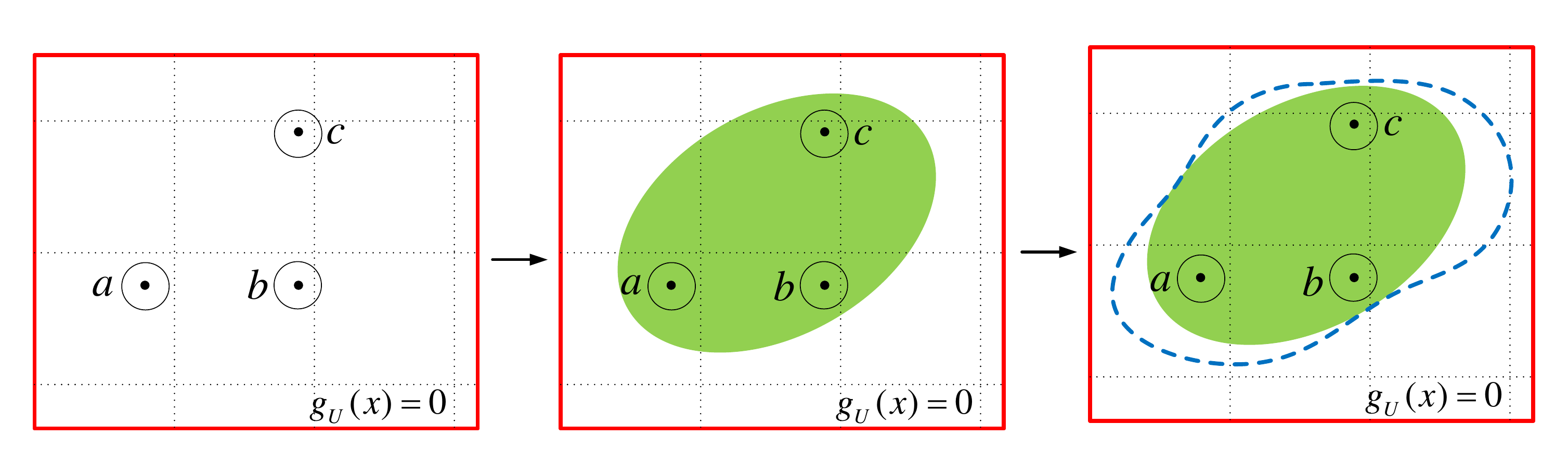}
	\caption{Demonstration of the iterative algorithm to estimate the largest ROS.}
	\label{fig_Iterative_Demo}
\end{figure}

Since the non-convexity is introduced by making the initial set as a variable, an iterative solution is proposed in \cite{zyc_hybrid_JCS_2017} starting by several guessed initial sets illustrated in Fig. \ref{fig_Iterative_Demo}. The principle of the proposed algorithm is to use the zero level set of a feasible barrier certificate as an initial condition and to search for a larger invariant set. Once feasible, this initial condition becomes the ROS due to the existence of corresponding invariant sets. But this algorithm does not provide information on how \emph{good} the estimation is.

A recent novel approach proposed in \cite{Henrion2014} uses occupation measures to formulate the reachability computation as an infinite-dimensional linear program. Its dual problem is formulated on the space of nonnegative continuous functions to compute the ROS shown in (\ref{eq_main}) 
\begin{problem}\label{thm_max_volume_om}
	\begin{subequations}
		\label{eq_main}
		\begin{align}
		&\inf_{B(x),\varOmega(x)} & &\int\limits_{X}\varOmega(x)d\lambda(x)\label{eq_main_1}\\
		&\text{s.t.} & & B(x)> 0\quad\forall x\in X_{U}\label{eq_main_2}\\
		& & &\dfrac{\partial B}{\partial x}f(x,d)\leq 0\quad\forall(x,d)\in X\times D\label{eq_main_3}\\
		& & &\varOmega(x)\geq B(x)+1\quad\forall x\in X\label{eq_main_4}\\
		& & & \varOmega(x)\geq 0\quad\forall x\in X\label{eq_main_5}
		\end{align}
	\end{subequations}
\end{problem}
The infimum is over $B\in C^{1}(X)$ and $\varOmega\in C(X)$. $\lambda$ denotes the Lebesgue measure. If the problem is feasible, the safety $f(x,d)$ with $d\in D$ is preserved and the zero level set of $\varOmega(x)-1$ converges below to $X_{I}^{*}$.\par

\begin{figure}[h]
	\centering
	\includegraphics[scale=0.6]{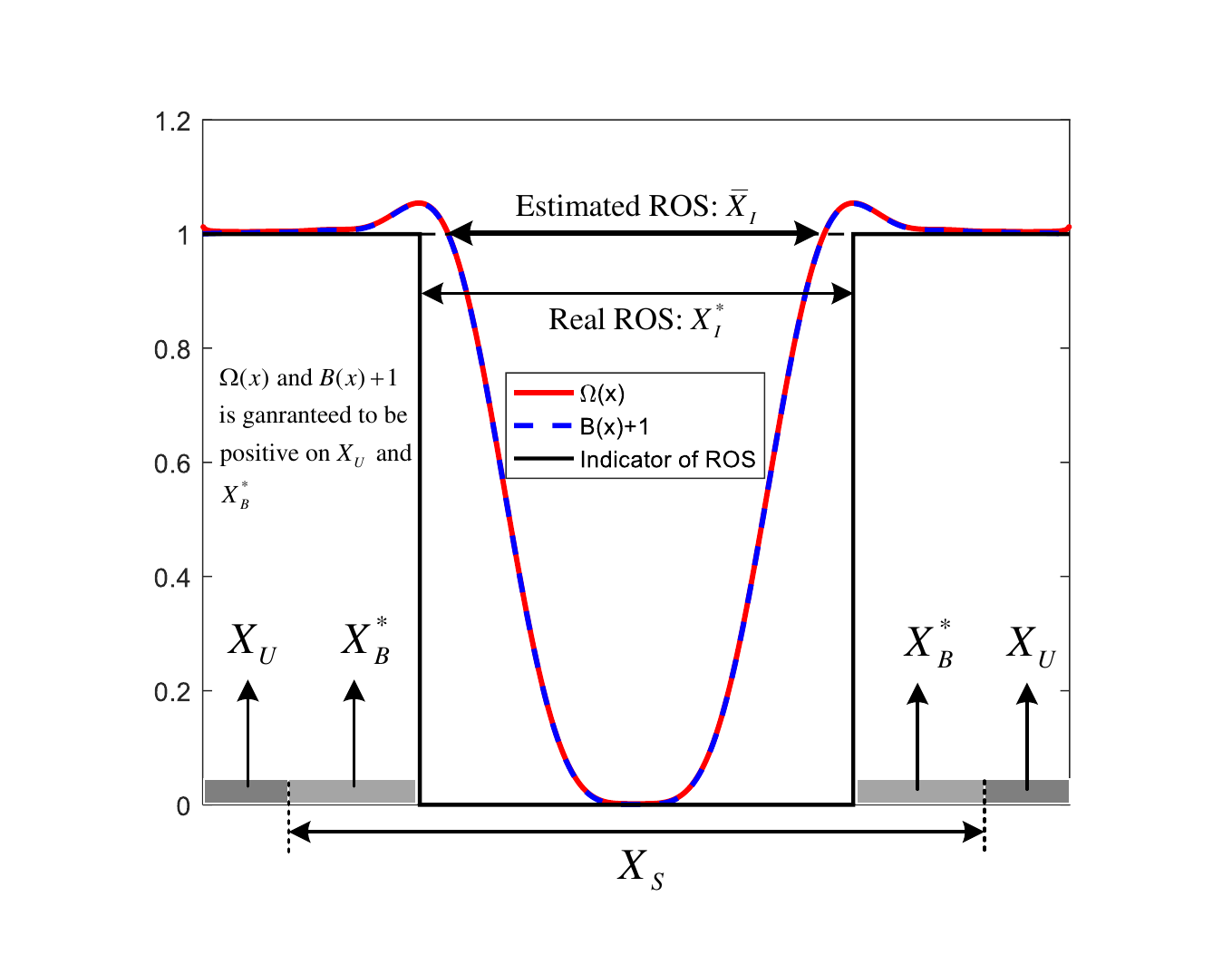}
	\caption{Geometry interpretation of proposed optimization problem for estimating the ROS. $\Omega(x)$ and $B(x)+1$ are guaranteed to be positive on $X_{U}$ and $X_{B}^{*}$.}
	\label{fig_Geometry}
\end{figure}

A strict mathematical proof is given in \cite{Henrion2014}, while a geometric interpretation is illustrated in \cite{zhang2018set}, which is briefly described as follows. Let any trajectory eventually ending up in the set $X_{U}$ at certain time $T$ be denoted as $\phi(T|x_{0})$. Based on the conditions of $B(\phi(T|x_{0}))>0$ in (\ref{eq_main_2}) and the passivity in (\ref{eq_main_3}), one can easily show $B(x_{0})>0$. Thus, (\ref{eq_main_2}) and (\ref{eq_main_3}) ensure that $B(x)>0$ for any $x\in X_{B}^{*}$ illustrated as a one dimensional case in Fig. \ref{fig_Geometry}. The conservatism lies in the fact that $B(x)>0$ for some $x\in X_{I}^{*}$, which overestimates the BRS (i.e., ${X}_{B}^{*}\subset\overline{X}_{B}$) and in turn underestimates the ROS (i.e., ${X}_{I}^{*}\supset\overline{X}_{I}$). Fortunately, this conservatism can be reduced by introducing a positive slack function $\varOmega(x)$ that is point-wise above the function $B(x)+1$ over the computation domain $X$. Assume the complement set of $X_{I}^{*}$ is represented by the indicator function $\delta_{X\setminus X_{I}^{*}}(x)$, i.e., a function is equal to one on $X\setminus X_{I}^{*}$ and 0 elsewhere. The key idea of the problem in (\ref{eq_main}) is that by minimizing the area of function $\varOmega(x)$ over the computation domain $X$, the function $B(x)+1$ will be forced to approach $\delta_{X\setminus X_{I}^{*}}(x)$ from above as shown in Fig. \ref{fig_Geometry}. Thus, the zero sublevel set of $\varOmega(x)-1$ is an inner approximation of $X_{I}^{*}$. Essentially, the problem in (\ref{eq_main}) is trying to approximate an indicator function using a polynomial. The conservatism of the estimate vanishes with increasing order of the polynomial.\par

\subsection{Positivity for Algorithmic Solutions}\label{sec_sub_BC_pos}
The key property for the barrier certificates is to enforce positivity or non-negativity (also denoted as semi-positivity) of functions over a given set $K \subseteq\mathbb{R}^{n}$ as
\begin{itemize}
	\item $p(x)$ is positive definite over a set $K$ if and only if for any $x \in K$, $p(x)>0$ 
	\item $p(x)$ is positive semi-definite over a set $K$ if and only if for any $x \in K$, $p(x)\geq 0$
\end{itemize}
Any such description is called a \emph{positivstellensatz} or \emph{nichtnegativstellensatz}, which ends with a combination of two German words \emph{stellen} (places) and \emph{satz} (theorem) \cite{Parrilo2000}. This is a very important problem, and a variety of efforts have been devoted to it. However, there is no general solution to prove the above property. To tackle the problem algorithmically, the classes of functions $p(x)$ have to be further restricted. A good compromise is achieved by considering the case of polynomial functions as every continuous function defined on a closed interval $[a,b]$ can be uniformly approximated as closely as desired by a polynomial function based on the Weierstrass approximation theorem.

Once confined to polynomial data, that is, the function $p(x)$ is polynomial and the set $K$ is defined by finitely many polynomial inequalities and equality constraints (denoted as semi-algebraic sets), the problem is solvable under certain cases. In 1900, Hilbert posted a list of 23 problems, the 17th of which was: Given a multivariate polynomial that takes only non-negative values over the reals, can it be represented as a SOS of rational functions \cite{reznick2000some}? The Hilbert's 17th problem was answered by Artin in 1927 \cite{Sassi2015}. But generally the positivity of polynomials is still under intensive studies, mainly being tackled from the algebraic geometry point of view \cite{powers2011positive}. From now on, we will focus on problems that are represented or approximated using polynomials. In this subsection, two main computation techniques are reviewed.

\subsubsection{SOS Representations}
\begin{definition}
	\label{thm_ch3_SOS}
	A polynomial $P(x)$ is a SOS if and only if there exist polynomials $p_1(x),\cdots,p_k(x)$ over $x$ such that $P(x)$ can be written as
	\begin{align}
	P(x)\equiv p_{1}^2(x)+\cdots+p_{k}^2(x)
	\end{align}
\end{definition}
We denote a SOS polynomial as $p\in\varSigma^{2}\left[ x\right]$. Any SOS polynomial is positive semi-definite over $\mathbb{R}^{n}$, while not every positive semi-definite polynomial is a SOS. A counter-example was provided by Motzkin known as the Motzkin polynomial shown as follows \cite{reznick2000some}
\begin{align}
M(x_1,x_2,x_3)= x_{1}^{4}x_{2}^{2} + x_{1}^{2}x_{2}^{4} - 3x_{1}^{2}x_{2}^{2}x_{3}^{2}+x_{3}^{6}
\end{align}
which is a non-negative degree 6 polynomial and is not a SOS.

For a positivstellensatz, it is sufficient to seek if $p$ is positive semi-definite over a semi-algebraic set $K$ represented as 
\begin{align}
K=\{x\in\mathbb{R}^{n}:g_{i}(x)\geq 0, g_{i}\in\mathbb{R}[x]\text{ for }i=1,\cdots,m\}
\end{align}
or written as $K:(g_1(x)\geq 0 \wedge\cdots\wedge g_m(x)\geq 0)$ for short. Then the following theorem can be used to verify the positivity \cite{Sassi2015}.
\begin{theorem}
	\label{thm_ch2_putinar_p}
	If a polynomial $p$ can be expressed as
	\begin{align}
	\label{eq_ch2_putinar}
	p\equiv q_0 + q_{1}g_{1} + \cdots + q_{m}g_{m}
	\end{align}
	for SOS polynomials $q_{0},q_{1},\cdots,q_{m}$, then $p$ is positive semi-definite over $K$.
\end{theorem}
Representing a polynomial in the form of (\ref{eq_ch2_putinar}) is denoted as the \emph{Putinar representation} \cite{kamyar2015polynomial}. In \cite{putinar1993positive} Putinar has proved that every polynomial that is strictly positive on $K$ has a Putinar representation. Thus, it is sufficient from computation point of view to search for a Putinar representation to provide the positivity certificate for a polynomial over a set. 

In most cases, $p_{i}(x)$ for $i=1,\cdots,k$ are constructed using the monomial basis under a bounded degree. Searching for appropriate coefficients such that $P(x)$ admits a sum of squares decomposition is denoted as the SOS programming (SOSP) and can be solved by relaxation to a semi-definite program (SDP) \cite{Parrilo2000,Parrilo2003}. Now Problem \ref{thm_max_volume_om} can be formally solved by the following problem.
\begin{problem}\label{thm_om_sos}
	\begin{subequations}\label{eq_sos}
		\begin{align}
		\inf_{B(x),\varOmega(x)} & \omega'l\\
		B(x)-\epsilon - \sigma_{1}(x)g_{U}(x)& \in \varSigma^{2}\left[ x\right]\\
		\begin{split}
		-\dfrac{\partial B}{\partial x}(x)f_{0}(x,d)-\sigma_{2}(x,d)g_{D}(d)\\-\sigma_{3}(x,d)g_{X}(x)&\in \varSigma^{2}\left[ x\right]\label{eq_sos_3}
		\end{split}\\
		\varOmega(x)-B(x)-1-\sigma_{4}(x)g_{X}(x) &\in \varSigma^{2}\left[ x\right]\\
		\varOmega(x)-\sigma_{5}(x)g_{X}(x) &\in \varSigma^{2}\left[ x\right]
		\end{align}
	\end{subequations}
\end{problem}
where $l$ is the vector of the moments of the Lebesgue measure over $X$ indexed in the same basis in which the polynomial $\varOmega(x)$ with coefficients $\omega$ is expressed. For example, for a two-dimensional case, if $\varOmega(x)=c_{1}x_{1}^{2}+c_{2}x_{1}x_{2}+c_{3}x_{2}^{2}$, then $\omega=[c_1,c_2,c_3]$ and $l=\int_{X}[x_{1}^{2},x_{1}x_{2},x_{2}^{2}]\text{d}x_{1}\text{d}x_{2}$.

Conversion of Problem \ref{thm_om_sos} to SDP has been implemented in solvers such as SOSTOOLS \cite{sostools} or the SOS module \cite{sos_yalmip} in YALMIP \cite{yalmip}. Then, the powerful SDP solvers like MOSEK can be employed \cite{mosek}.

\subsubsection{Linear Representations}
As an alternative to the SOS representation, another class of linear representations involves the expression of the target polynomial to be proven non-negative over the set $K$ as a linear combination of polynomials that are known to be nonnegative over the set $K$. This approach reduces the polynomial positivity problem to a linear program (LP) \cite{kamyar2015polynomial}\cite{Sassi2015}. Then the so-called \emph{Handelman representations} are employed to ensure the non-negativity of a polynomial form over a region. Let $K$ be defined as a semi-algebraic set again: $K=\{x\in\mathbb{R}^{n}:p_{j}(x)\geq 0,j=1,2,\cdots,m\}$. Denote the set of polynomials $P$ as $\{p_{1},p_{2},...,p_{m}\}$. This approach writes the given polynomial $p(x)$ as a conic combination of products of the constraints defining $K$, i.e., $p(x)=\lambda_{f}f$, where $\lambda_{f}\in\mathbb{R}^{+}$ are the coefficients, $D$ is the bounded degree and $f$ belongs to the following set
\begin{equation}
f\in\mathcal{P}(P,D)=\{p_{1}^{n_{1}}p_{2}^{n_{2}}\cdots p_{m}^{n_{m}}:n_{j}\leq D,j=1,2,\cdots,m\}
\end{equation}
If the semi-algebraic set reduces into a polyhedron, that is, $p_{j}(x)=a_{j}x-b_{j}$, then the following conclusion known as the Handelman's Theorem provides a useful LP relaxation for proving polynomial positivity \cite{Handelman1988}.
\begin{theorem}[Handelman]
	\label{thm_ch3_handelman}
	If $p(x)$ is strictly positive over a compact polyhedron $K$, there exists a degree bound $D>0$ such that
	\begin{align}
	\label{eq_ch3_handelman}
	p(x)=\sum\lambda_{f}f \text{ for } \lambda_{f}\geq 0 \text{ and } f\in\mathcal{P}(P,D)
	\end{align}
\end{theorem}
An example in \cite{Sassi2015} is presented here for better illustration. Consider the polynomial $p(x_1,x_2)=-2x_{1}^{3} + 6x_{1}^{2}x_{2} + 7x_{1}^{2} - 6x_{1}x_{2}^2 -14x_{1}x_{2} + 2x_{2}^{3} + 7x_{2}^{2} - 9$ and the set $K:(x_1 - x_2 - 3\geq 0 \wedge x_2 - x_1 - 1\geq 0)$. Then, the positivity of $p$ over $K$ can be proved by representing $p$ as follows
\begin{align}
p(x_1,x_2)=\lambda_{1}f_{1}^{2}f_{2}+3f_{1}f_{2}
\end{align}
where $f_{1}=x_1 - x_2 - 3$, $f_{2}=x_2 - x_1 - 1\geq 0$, $\lambda_{1}=2$ and $\lambda_{2}=3$.

The general procedure is described as follows \cite{Sassi2015}:
\begin{enumerate}
	\item Choose a degree limit $D$ and construct all terms in $\mathcal{P}(P,D)$, where $P=\{p_{1},p_{2},...,p_{m}\}$ are the lines defining polyhedron $K$.
	\item Let $p(x)=\sum_{f\in\mathcal{P}(P,D)}\lambda_{f}f$ for unknown multipliers $\lambda_{f}\geq 0$.
	\item Equate coefficients on both sides (the given polynomial and the Handelman representation) to obtain a set of linear inequality constraints involving $\lambda_{f}$. 
	\item Use a LP solver to solve these constraints. If feasible, the results yields a proof that $p(x)$ is positive semi-definite over $K$.
\end{enumerate}

Handelman's Theorem results in a LP, and thus reduces the computation burden. However, since the multipliers $\lambda_{f}$ are real numbers instead of SOS polynomials in Putinar representation, it admits a less chance to find a Handelman representation, leaving the problem inconclusive.

\subsubsection{An Illustrative Example}
\begin{figure}[h]
	\centering
	\includegraphics[width=3.0 in]{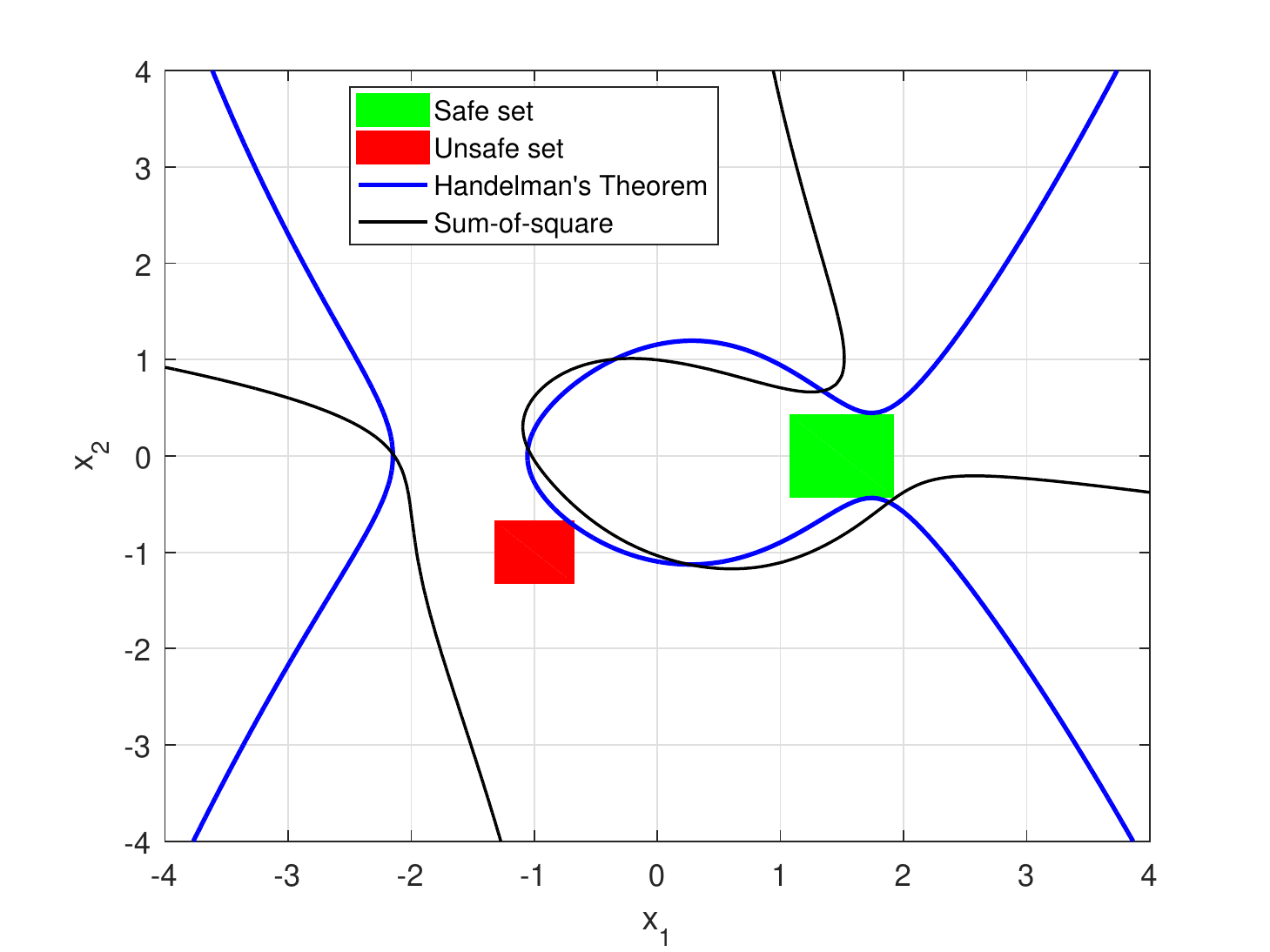}
	\caption{Safety verification using the Handelman representation.}
	\label{fig_Handelman}
\end{figure}

We employ the example in \cite{Prajna2007a} to illustrate these two representation by solving Theorem \ref{thm_barrier_a} as a precursor. Similar attempt is made in \cite{Yang2016b} as well. Consider the following system
\begin{align}
\left[\begin{array}{c}\dot{x}_{1}\\\dot{x}_{2}\end{array}\right] =
\left[ \begin{array}{c} 
x_{2}\\
-x_{1}+\frac{1}{3}x_{1}^{3}-x_{2}
\end{array} \right]
\end{align}
The original sets are defined as: $X=\mathbb{R}^{2}$, $X_{I}=\{x\in\mathbb{R}^{2}:(x_{1}-1.5)^{2}+x_{2}^{2}\leq 0.25\}$, $X_{U}=\{x\in\mathbb{R}^{2}:(x_{1}+1)^{2}+(x_{2}+1)^{2}\leq 0.16\}$. To employ the Handelman's Theorem, they are modified to be polyhedrons as shown in Fig. \ref{fig_Handelman}. The barrier certificate computed using the Handelman's Theorem is plotted as the blue curve, while the one obtained by SOSP is plotted as the dark curve. As seen, although the barrier certificates are different, both approaches successfully verify the safety of the system. 

\subsection{Power System Applications}\label{sec_sub_BC_app}
The initial application of barrier certificates appeared in \cite{Wisniewski2013} and \cite{Laurijsse2014a}. The barrier certificate methodology is employed to design the safety supervisor such that the wind turbines can be shutdown timely in emergent conditions. Voltage constraint satisfaction under variable distributed generation and time-varying consumption is verified in \cite{Pedersen2016}. In \cite{zyc_hybrid_JCS_2017,zhang2018set}, a safety supervisory control is designed to timely activate the inertia emulation functions within a wind turbine generator such that the system frequency is adequate with respect to a given worst case. In \cite{kundu2019distributed}, a control policy is  designed and certified using barrier certificates such that the voltage limits during transients are respected under generated active and reactive power setpoints. Closely related works are the stability analysis based on Lyapunov functions \cite{anghel2013algorithmic,kundu2015stability,mishra2017stability,mishra2019transient,josz2019transient}.

One advantage of passivity-based methods compared with the set operation-based methods is that the certificate is a function of system states. As analyzable and quantifiable, the certificates can be readily employed as a supervisory control for multi-mode control systems such as grid-interactive converters.  This supervisory control can not only generate switching commands, but also provide real-time margin for a critical safe switching. The works in \cite{Wisniewski2013,Laurijsse2014a,zyc_hybrid_JCS_2017,zhang2018set} have taken this advantages.

\section{Benchmark Example}
\begin{figure}[!h]
	\centering
	\includegraphics[width=3 in]{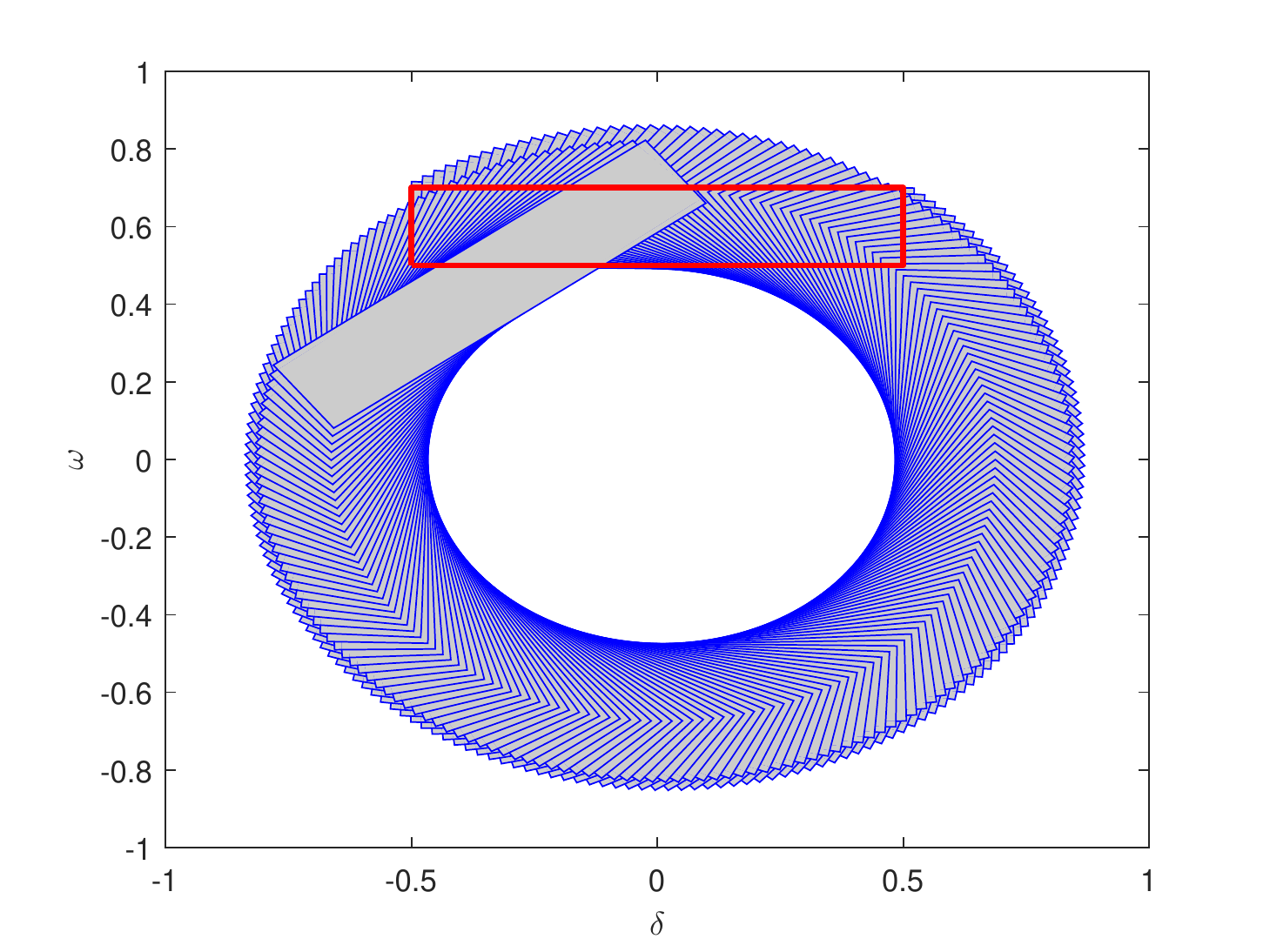}
	\caption{Backward reachable set computation using zonotopes for 1 s in reverse time. $x_{1}$ is the rotor angle and $x_{2}$ is the machine speed.}
	\label{fig_zonotope}
\end{figure}

\begin{figure}[h]
	\centering
	\includegraphics[width=3 in]{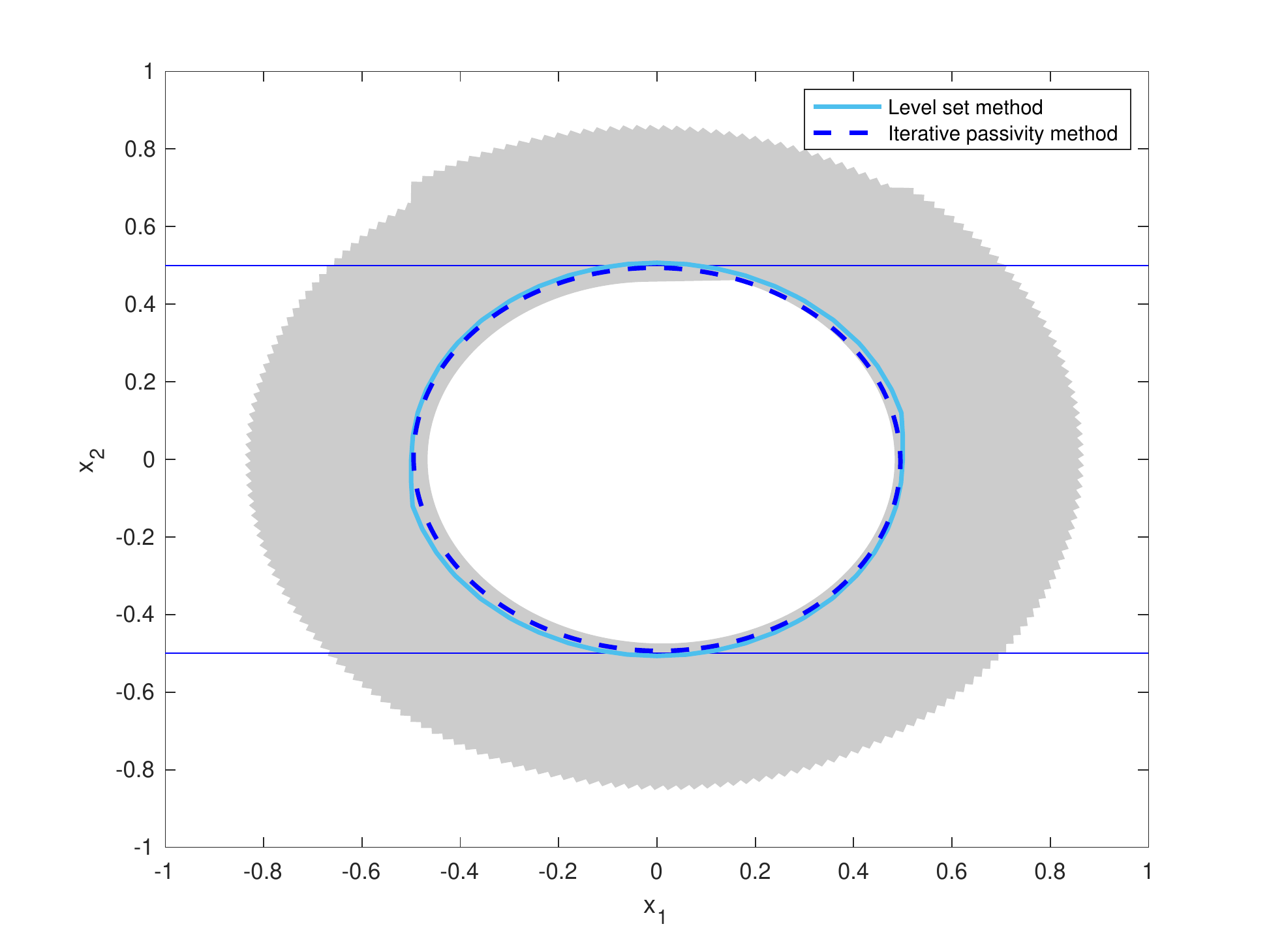}
	\caption{ROS computed by the level set method and iterative algorithm in Fig. \ref{fig_Iterative_Demo} and compared with the backward reachable set of the unsafe set using zonotope representations.}
	\label{fig_ROS_Comparison}
\end{figure}

\begin{figure}[h]
	\centering
	\includegraphics[width=3 in]{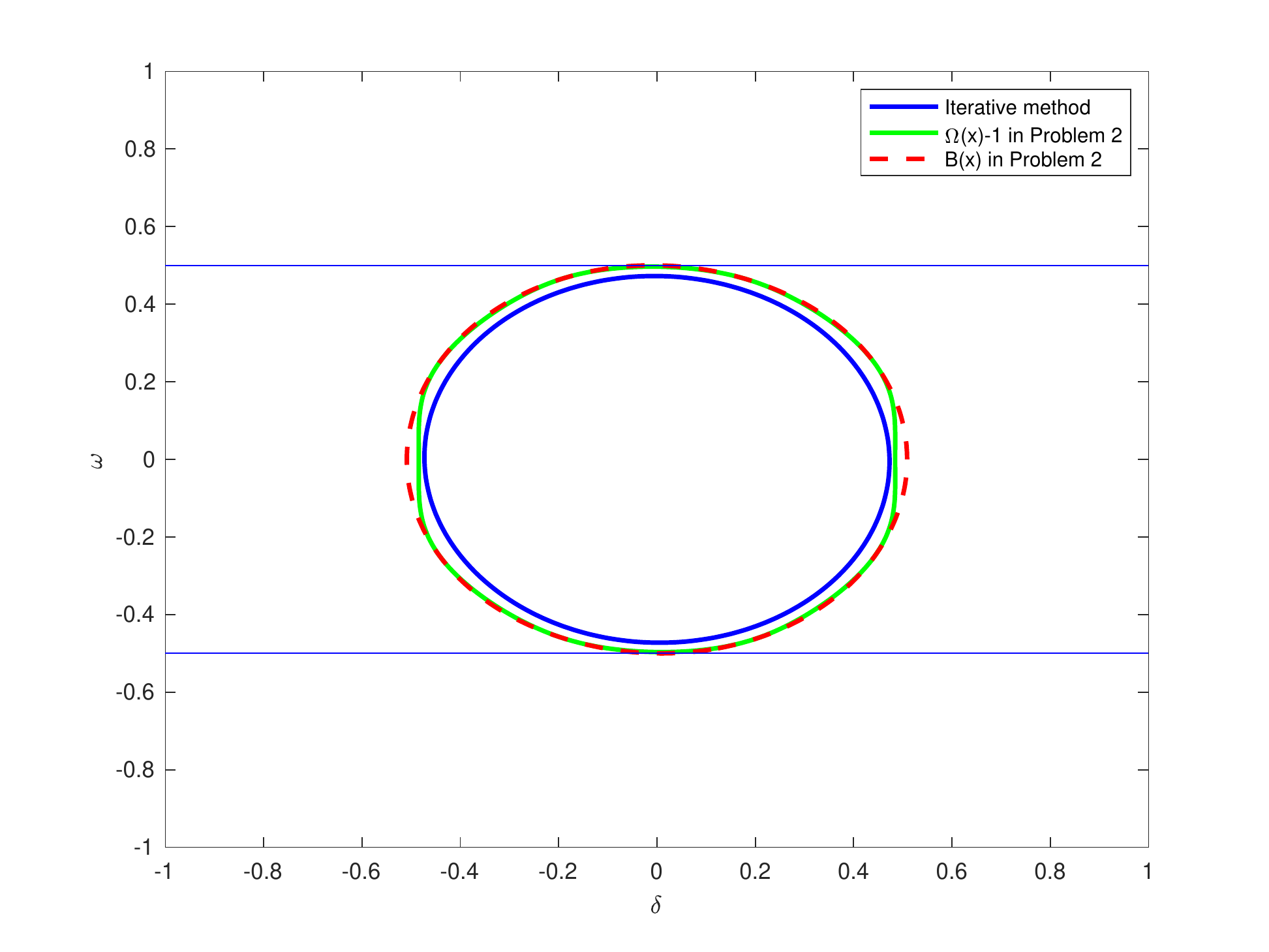}
	\caption{ROS computed by the algorithm in Fig. \ref{fig_Iterative_Demo} and Problem \ref{thm_max_volume_om}.}
	\label{fig_ch4_ROS_Comparison2}
\end{figure}

To further demonstrate the approaches, a simple example is illustrated as a benchmark. Lagrangian methods, Eulerian Method and passivity-based methods are compared showing highly consistent results. Consider the linearized single-machine infinite-bus system as follows
\begin{align}
\left[\begin{array}{c}\Delta\dot{\delta}\\\Delta\dot{\omega}\end{array}\right] =
\left[ \begin{array}{cc} 
0 & 6.2833\\
-6.2696 & -0.1429
\end{array} \right]
\left[\begin{array}{c}\Delta\delta\\\Delta\omega\end{array}\right]
\end{align}
Define the safety specification as $X_U=\{[\delta,\omega]^{T}: -0.5\leq\omega\leq 0.5\}$. First, the zonotope-based set operating method is applied in backward to find the largest backward reachable set of the unsafe set. Define an unsafe set as the red box shown in Fig. \ref{fig_zonotope} and propagate this set in reverse time.
If the computation is long enough, then an invariant set in the middle of the backward reachable set of the unsafe set is obtained, which is actually the ROS. The ROSs computed by the level set method and the iterative algorithm in Fig. \ref{fig_Iterative_Demo} are shown in Fig. \ref{fig_ROS_Comparison} together with the backward reachable set via the zonotope method. The three results are in accordance with each other, and the backward reachability interpretation of the largest ROS is verified. 
The results obtained by the algorithm in Fig. \ref{fig_Iterative_Demo} and Problem \ref{thm_max_volume_om} are compared in Fig. \ref{fig_ch4_ROS_Comparison2}.
In this simple case, the two results are consistent. The zero level set of $B(x)$ solved by Problem \ref{thm_max_volume_om} is enlarged by $\varOmega(x)-1$ as much as possible to the largest ROS under the fixed highest degree. With increasing dimensions of the system, higher degrees may need to obtain a convergent result from Problem \ref{thm_max_volume_om}. Limited by the computation complexity, Problem \ref{thm_max_volume_om} sometimes fails to converge. The algorithm in Fig. \ref{fig_Iterative_Demo} can always provide certain results, however, with unknown conservatism.

\section{Conclusion}\label{sec_con}
In this paper, set-theoretic methods for power system safety verification and control are reviewed. The methods are categorized into set operation-based and passivity-based methods according to their underlying mathematical principle. In general, set operating-based methods are computationally more efficient and applicable to higher-order systems. On the other hand, passivity-based methods provide semi-analytical representations of reachable sets and can be readily deployed for multi-mode control systems. A benchmark example is given. The ROS is computed via different methods, resulting in high consistency. The reviewed methods provide vivid solutions to handle unknown-but-bounded uncertainty in power system operations.

\section{Future Research}
Generally speaking, however, scalability of set-theoretic methods is the most challenging factor that prohibit it from power system application as realistic power networks are significantly large-scale. The future research efforts should be dedicated to the improvement of the scalability of these approaches. For Lagrangian methods, one direction is to decompose the system and perform distributed and parallel reachable set computation \cite{li2018networked}. The decomposition will need further investigating based on the feature of the underlying system. On the other hand, algorithms that can incorporate model reduction techniques, such as Krylov subspace approximation methods \cite{althoff2019reachability}, into the reachability computation would be promising. More importantly, these algorithms should be able to accurately approximate all or many states of the systems, which are usually the studying objectives. For Eulerian methods, since it relies on standard solving procedure of partial differential equations, the scalability can be improved mainly from modeling and order reduction perspective. Nevertheless, recent research has revealed its close connection to the passivity-based methods \cite{Henrion2014,zhang2018set}. Strict mathematical proof to demonstrate their equivalence under certain assumptions is worth studying. For passivity-based approaches, polynomial selection to admit lighter weighted program has been extensively studied. Diagonally dominant sum of squares (DSOS) and scaled diagonally dominant sum of squares (SDSOS) optimization as linear programming and second-order cone programming--based alternatives to sum of squares optimization that allow one to trade off computation time with solution quality. These are optimization problems over certain subsets of sum of squares polynomials \cite{ahmadi2019dsos}. In addition, applying different positivity certificates, such as Handelman's representation \cite{Sassi2015} and Krivine-Stengle’s certificate \cite{lasserre2013lagrangian}, will lead to linear program. All these aforementioned approaches have not been employed in research for large-scale power systems. Moreover, the impact of the choice of the polynomial basis (e.g., Chebyshev, trigonometric or power) on the quality of the solution of the SDP relaxations deserves further investigation for a better understanding \cite{henrion2009approximate}.


\bibliography{Ref_Set}  
\bibliographystyle{IEEEtran}

\end{document}